\documentclass[final,3p,times,twocolumn]{elsarticle}
\usepackage{epstopdf}

\usepackage[colorlinks,citecolor=blue,linkcolor=blue,anchorcolor=blue,filecolor=blue,urlcolor=blue]{hyperref}
\usepackage{amssymb,amsmath}
\usepackage{CJKutf8}
\biboptions{sort&compress}

\journal{Physics Letters B}

\begin{document}
\begin{CJK*}{UTF8}{gbsn}
\begin{frontmatter}

\title{Production of proton-rich nuclei in the vicinity of $^{100}$Sn via multinucleon transfer reactions}
\author[UCAS]{Zhenji Wu ~(吴振基)}
\author[UCAS,ITP]{Lu Guo ~(郭璐)\corref{cor1}}
  \ead{luguo@ucas.ac.cn}
\cortext[cor1]{Corresponding author.}
\author[IMP,UCAS]{Zhong Liu ~(刘忠)}
\author[UCAS,HNU]{Guangxiong Peng ~(彭光雄)}

\address[UCAS]{School of Nuclear Science and Technology, University of Chinese Academy of Sciences, Beijing 100049, China}
\address[ITP]{CAS Key Laboratory of Theoretical Physics, Institute of Theoretical Physics, Chinese Academy of Sciences, Beijing 100190, China}
\address[IMP]{Institute of Modern Physics, Chinese Academy of Sciences, Lanzhou 730000, China}
\address[HNU]{Synergetic Innovation Center for Quantum Effects and Application, Hunan Normal University, Changsha 410081, China}

\begin{abstract}
The production of new proton-rich nuclei in the vicinity of $^{100}$Sn is investigated via multinucleon transfer reactions within the framework
of microscopic time-dependent Hartree-Fock (TDHF) and statistical model \texttt{GEMINI++}. The TDHF+GEMINI method has demonstrated
the reliable description in the multinucleon transfer dynamics and the agreement between theoretical results
and experimental data is quite satisfactory in the observed transfer reactions. We reveal the production cross sections of proton-rich nuclei in $^{100}$Sn region via multinucleon transfer reactions to be
several orders of magnitude higher than those measured via fusion-evaporation and projectile fragmentation experiments.
About 19 new proton-rich isotopes with cross sections of larger than 1 nb are predicted to be produced in multinucleon transfer reaction of $^{58}$Ni with $^{112}$Sn.
The reaction mechanisms are discussed to lead the experimental production of these previously unreported nuclei. Multinucleon transfer reactions provide a fascinating possibility to reach the proton drip-line in $^{100}$Sn region and beyond.
\end{abstract}

\begin{keyword}
Multinucleon transfer reaction \sep
Quantum shell effect \sep
Proton-rich nuclei \sep
Time-dependent Hartree-Fock


\end{keyword}
\end{frontmatter}
\end{CJK*}
\section{INTRODUCTION}
The proton-rich nuclei in the vicinity of $^{100}$Sn, the heaviest self-conjugate doubly magic nucleus discovered in the chart of nuclei~\cite{Celikovic2016_PRL116-162501}, exhibit unique nuclear structure features and play crucial role in the astrophysical rapid proton (rp) capture process. This region has been the subject of intense experimental and theoretical studies in the last two decades, see~\cite{Faestermann2013_PPNP69-85} and references therein.
Recent $\beta$-decay spectroscopy of $^{100}$Sn~\cite{Hinke2012_Nature486-341,Lubos2019_PRL122-222502}, establishing the superallowed nature of the $^{100}$Sn
Gamow-Teller decay and together with the isomeric decay spectroscopy below $^{100}$Sn~\cite{NaraSingh2011_PRL107-172502,Davies2019_PRC99-021302}, provide strong evidence for the robustness of the $N=Z=50$ shell closures. Due to the doubly magic nature of $^{100}$Sn, an island of enhanced $\alpha$ emitters above $^{100}$Sn~\cite{Liddick2006_PRL97-082501,Darby2010_PRL105-162502} is expected and preliminary evidence for the super-allowed $\alpha$ decay to $^{100}$Sn has been reported recently~\cite{Auranen2018_PRL121-182501}.
This fascinating region is crossing the proton drip-line and proton-unbound nuclei have been identified both below and above $^{100}$Sn.

Despite continuous efforts, progress in the experimental investigation in this region has been slow over the years, because of the very low production yield of these very proton-rich nuclei. Nuclei in this region have been produced by projectile fragmentation of intermediate to high-energy heavy-ion beams and fusion-evaporation reactions between low-energy ions. The most proton-rich nuclei below $^{100}$Sn have mainly been produced in projectile fragmentation reactions~\cite{Celikovic2016_PRL116-162501}. For those above $^{100}$Sn, fusion-evaporation reactions are still advantageous, e.g., the heaviest self-conjugate isotope $^{108}$Xe has been identified via fusion-evaporation reaction and the production cross section has been measured to be less than 1 nb~\cite{Auranen2018_PRL121-182501}.
The prospects for extending experimental research towards the more exotic nuclei in this region are severely limited by the beam intensities available at current facilities. It is of great importance to explore an alternative reaction mechanism for the production of proton-rich nuclei in this region.

In recent years multinucleon transfer (MNT) reaction occurring in low-energy collisions has been considered as a promising method of enormous potential to produce new unstable nuclei, especially on the neutron-rich side, which are difficult to be produced by other reactions. For example, the hard-to-reach neutron-rich nuclei around $N=126$ have been successfully produced via MNT reactions~\cite{Barrett2015_PRC91-064615, Vogt2015_PRC92-024619, Watanabe2015_PRL115-172503}, and the measured cross sections are several orders of magnitude larger than those from projectile fragmentation of high-energy heavy-ion beams~\cite{Watanabe2015_PRL115-172503}.
The MNT experiment with actinide projectile and target nuclei indicated the possible production of superheavy nuclei with atomic numbers as high as 116~\cite{Wuenschel2018_PRC97-064602}, which provides an alternative pathway for the production of superheavy elements.

In present work, we propose to produce the proton-rich nuclei in the vicinity of $^{100}$Sn via MNT reactions in low-energy
collision of $^{58}$Ni with $^{112}$Sn, and show its advantage over the conventional production through fusion-evaporation 
and projectile fragmentation reactions.
The idea is to take advantage of transfer mechanisms and the stabilizing effect of the closed proton shells in projectile and target nuclei.
The crucial transfer mechanisms determining the mass and charge of reaction products are charge equilibrium and neck evolution in transfer process.
Our recent study of MNT dynamics~\cite{Wu2019_PRC100-014612} revealed that neck formation
and abruption are
expected to be dominant at the relatively small impact parameters leading to both neutrons and protons transferred in the same direction, while
at large impact parameters the predominant mechanism of charge equilibrium leads to the opposite transfer of protons and neutrons.
Concurrently, the shell effect plays a very important role in the transfer process.
These reaction mechanisms have been automatically taken into account in our microscopic approaches~\cite{Wu2019_PRC100-014612}.
To compare the production of nuclei around $^{100}$Sn in different reaction systems, 
MNT reactions of $^{58}$Ni with $^{106}$Cd and $^{124}$Xe are also investigated in this work.

Various theoretical models~\cite{Adamian2010_PRC81-057602,Wang2012_PRC85-041601,Zagrebaev2013_PRC87-034608,Wen2013_PRL111-012501,Wen2014_PRC90-054613,
Zhao2016_PRC94-024601,Bao2016_PRC93-044615,Feng2017_PRC95-024615,Zhu2017_PLB767-437,Li2018_PLB776-278,Xu2019_CPC43-064105} have been developed to describe the multinucleon transfer process. In this work, we combine the microscopic time-dependent Hartree-Fock (TDHF) approach~\cite{Simenel2012_EPJA48-152,Nakatsukasa2016_RMP88-045004,Simenel2018_PPNP103-19,Stevenson2019_PPNP104-142,Sekizawa2019_FP7-20} with the state-of-the-art statistical model \texttt{GEMINI++}~\cite{GEMINI++_code,Charity2010_PRC82-014610} to describe such reaction process.
The initial multinucleon transfer stage is depicted with TDHF approach, and
particle-number projection technique is applied on TDHF wave functions to extract
the transfer probability for each transfer channel. The subsequent deexcitation process including the emission of light particles and fission of heavy fragments is described with the statistical model \texttt{GEMINI++}.
Very recently, the combined approach TDHF+GEMINI~\cite{Sekizawa2017_PRC96-041601,Sekizawa2017_PRC96-014615,Jiang2018_CPC42-104105,Wu2019_PRC100-014612,Wu2020_SCPMA63-242021,Jiang2020_PRC101-014604} has been applied in multinucleon transfer reactions and reasonably reproduced the experimental cross sections for the final products.
In addition, the combined method also well accounts for the experimental observation of isotopic dependence of fusion-evaporation cross sections for the
production of superheavy elements~\cite{Guo2018_PRC98-064609}. These results are very remarkable since reaction calculations do not require additional
parameters related to the reaction dynamics.

The article is organized as follows. The microscopic TDHF approach, particle-number projection, and statistical model \texttt{GEMINI++} are 
briefly reviewed in Sec.~\ref{theory}. Section~\ref{results} presents the multinucleon transfer dynamics for the production of new proton-rich nuclei 
in $^{100}$Sn region.
A brief summary is given in Sec.~\ref{summary}.

\section{Theoretical approaches}
\label{theory}
TDHF is an approximate approach to solve the quantum many-body problem and has wide applications in low-energy heavy-ion collisions,
see recent publications~\cite{Umar2014_PRC89-034611,Dai2014_PRC90-044609,Dai2014_SciChinaPMA57-1618,	Wang2016_PLB760-236,Umar2016_PRC94-024605,Stevenson2016_PRC93-054617,Simenel2017_PRC95-031601,
Yu2017_SciChinaPMA60-092011,Umar2017_PRC96-024625,Guo2018_PRC98-064609,Guo2018_PLB782-401,Guo2018_PRC98-064607,Scamps2018_Nature564-382,Sekizawa2019_PRC99-051602,Li2019_SciChinaPMA62-122011,Godbey2019_PRC100-054612,Ayik2019_PRC100-044614}.
The important dynamics in multinucleon transfer process, such as deformation evolution of nuclear system, shell and quantum effects, energy dissipation, and nucleon exchanges
has been incorporated in TDHF, which are essential for the manifestation of reaction mechanism and collision dynamics.

In TDHF approach, the many-body wave function $\Phi$ is assumed to be a Slater determinant
\begin{equation}\label{eq:1}
\Phi\left(\mathbf{r}_1,\cdots,\mathbf{r}_A,t\right) = \frac{1}{\sqrt{A!}}\text{det}\{\phi_\lambda\left(\mathbf{r}_i,t\right)\}, \ \lambda=1,\cdots,A 
\end{equation}
in the dynamic process.     
The time-dependent single-particle states $\phi_\lambda\left(\mathbf{r},t\right)$ satisfy the TDHF equation
\begin{equation}
i\hbar\frac{\partial}{\partial t}\phi_{\mathrm{\lambda}}(\mathbf{r},
t)=h\phi_{\mathrm{\lambda}}(\mathbf{r}, t),
\end{equation}
where $h$ denotes the single-particle Hamiltonian.

To extract the transfer probability $P_{Z,N}$ for each transfer channel, the particle-number projection method~\cite{Simenel2010_PRL105-192701} is applied to project the TDHF wave functions into the eigenstates with good proton and neutron number $Z, N$. The cross section of primary product is obtained by  integrating $P_{Z,N}$  over impact parameter $b$  
\begin{equation}
\sigma_{Z,N}(E)=2\pi\int_{b_{\text{min}}}^{b_{\text{cut}}}b P_{Z,N}(b,E)\text{d}b,
\label{Eq:cs_pri}
\end{equation}
where  $b_{\text{cut}}$ and $b_{\text{min}}$ are the  upper and lower limits of the impact parameter at which the binary reaction occurs.

The excited primary products will undergo the deexcitation process including the emission of light particles and fission of heavy fragments. 
This process is described with the Monte-Carlo-based statistical model \texttt{GEMINI++}.
The decay probability $P_\text{decay}(Z,N;Z^\prime,N^\prime)$ from primary product with $Z$ protons and $N$ neutrons to final product with $Z^\prime$ protons and  $N^\prime$ neutrons is evaluated by
\begin{equation}
P_\text{decay}(Z,N;Z^\prime,N^\prime)=\frac{M_{Z^\prime,N^\prime}}{M_{\textrm{event}}},
\label{Eq:decay_pro}
\end{equation}
where $M_{Z^\prime,N^\prime}$ is the event number of final products with $(Z^\prime, N^\prime)$ in the total $M_{\textrm{event}}$ simulations.
The production probability for final product with $Z^\prime$ protons and $N^{\prime}$ neutrons is the product of transfer and decay probabilities
\begin{equation}
	P^{(\text{final})}_{Z^\prime,N^\prime}(b,E)=\sum_{Z\geq Z^\prime}\sum_{N\geq N^\prime}P_{Z,N}(b,E)\times P_{\text{decay}}(Z,N;Z^\prime,N^\prime).
\end{equation}
The cross section for final fragments is calculated by integrating $P^{(\text{final})}_{Z^\prime,N^\prime}$
\begin{equation}
	\sigma^{(\text{final})}_{Z^\prime,N^\prime}(E)=2\pi\int_{b_{\text{min}}}^{b_{\text{cut}}} b P^{(\text{final})}_{Z^\prime,N^\prime}(b,E)\text{d}b.
\end{equation}

We solve TDHF equation in a symmetry-unrestricted three dimensional grid $56\times 24 \times 46$ with a grid spacing 1 fm.
The reaction is in $x$-$z$ plane and the collision axis is along $x$-axis.
We utilize Skyrme effective interaction SLy5~\cite{Chabanat1998_NPA635-231_NPA643-441}, in which all the time-even and time-odd terms in the mean-field Hamiltonian are included
in our code~\cite{Yu2017_SciChinaPMA60-092011,Guo2018_PRC98-064609,Guo2018_PLB782-401,Guo2018_PRC98-064607,Wu2019_PRC100-014612,Wu2020_SCPMA63-242021}. The static calculations are performed on three-dimensional grid $24\times 24 \times 24$ with a grid spacing 1 fm.
We find that the HF ground states of $^{58}$Ni and $^{106}$Cd show prolate deformations, while $^{112}$Sn and $^{124}$Xe present triaxially deformed
ground states.
It should be noted that with the inclusion of pairing correlations, the ground-state deformations of the projectile and target nuclei exhibit spherical symmetry for $^{58}$Ni and $^{112}$Sn and axial symmetry for $^{106}$Cd, while a coexistence of triaxially deformed ground state and prolately deformed local minimum with a small energy difference of 140 KeV is observed for $^{124}$Xe. The ground-state deformation effect of pairing correlations may have an impact on the transfer dynamics and pairing could also affect the reaction mechanisms in the dynamical evolution. However, the effect of pairing correlations in multinucleon transfer reaction is still an open question, because the inclusion of pairing requires much more computational cost in the microscopic simulation of collision dynamics. In this work, we concentrate on the systematic
TDHF studies and leave the inclusion of pairing correlation for future works.

\section{Results and discussions}
\label{results}

Since both projectile and target are deformed, the complete reaction dynamics should be done by the proper average over all the deformation orientations.
However, since the microscopic TDHF computation is very time-consuming, we calculate the TDHF dynamics in two extreme collisions of tip and side,
in which both projectile and target nuclei are in the tip and side orientations. The so-called tip (side) orientation for the deformed nucleus indicates
that the long (short) axis of nucleus is initially set along the collision axis. For the collisions of the proposed reactions,
we extract the fusion barrier from the nucleus-nucleus potential by using the frozen Hartree-Fock (FHF) method, in which both projectile and target nuclei are assumed to
keep their ground-state densities~\cite{Yu2017_SciChinaPMA60-092011,Guo2018_PRC98-064609,Guo2018_PLB782-401,Guo2018_PRC98-064607,Wu2019_PRC100-014612,Wu2020_SCPMA63-242021}.
The method and programming details to calculate the nucleus-nucleus potential by using the FHF approximation can be found in Ref.~\cite{Stevenson2020_IOPSciNotes1-025201}.
The fusion barrier is found to be $V_\text{B}$=151.89, 158.40, 163.96 MeV for the reactions of $^{58}$Ni with $^{106}$Cd, $^{112}$Sn, and $^{124}$Xe, respectively,
which becomes higher with the increase of mass number of target nuclei as expected.

\begin{figure}
\includegraphics[width=0.48\textwidth]{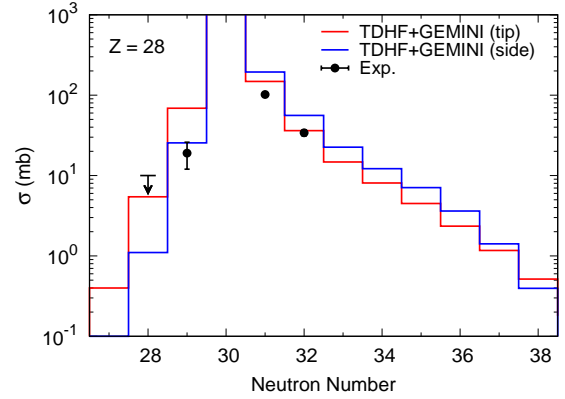}\\
\caption{(Color online) Production cross sections of Ni isotopes in multinucleon transfer reaction $^{58}$Ni+$^{112}$Sn. The TDHF+GEMINI results
for the tip and side collisions at $E_\text{c.m.}=221.77$ MeV (corresponding to $1.4V_\text{B}$) are denoted by the red and blue histograms,
respectively. The experimental data at $E_\text{c.m.}=217$ MeV~\cite{Berg1988_PRC37-178} is included for comparison.}
\label{Fig:experiment}
\end{figure}

\begin{figure}
\includegraphics[width=0.48\textwidth]{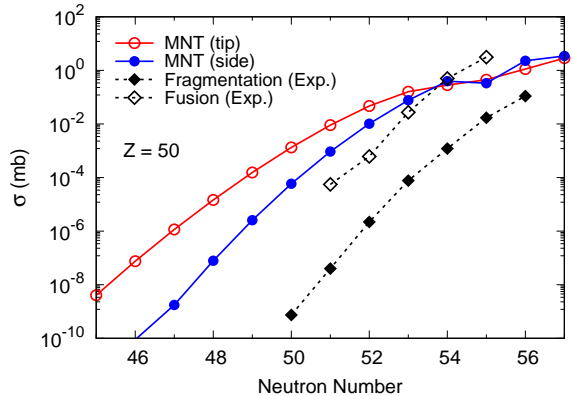}\\
\caption{(Color online) Production cross sections for proton-rich Sn isotopes using multinucleon transfer (MNT), projectile fragmentation, and fusion-evaporation reactions.
The MNT reaction $^{58}$Ni+$^{112}$Sn is performed with TDHF+GEMINI approach at center-of-mass energy of $1.4V_\text{B}$ for the tip (open circles) and side (solid circles) collisions, respectively. The measured cross sections using projectile fragmentation~\cite{Suzuki2013_NIMB317-756} (solid diamonds) and fusion-evaporation 
reactions~\cite{Karny2005_EPJA25-135} (open diamonds) are included for comparison.}
\label{Fig:orientation}
\end{figure}

We first compare our calculated cross sections with the available experimental data. 
The experiment performed in Argonne National Laboratory has measured the
transfer cross sections in $^{58}$Ni+$^{112}$Sn at center-of-mass (c.m.) energy of 217 MeV~\cite{Berg1988_PRC37-178}, which are shown by solid circles with errorbars in Fig.~\ref{Fig:experiment}.
The symbol of down arrow indicates the upper limit of experimental measurement.
The cross sections of Ni isotopes from TDHF+GEMINI calculations
at $E_\text{c.m.}=221.77$ MeV (corresponding to $1.4V_\text{B}$) for the tip and side collisions
are denoted by the red and blue histograms, respectively.
We find that the production cross sections obtained from TDHF+GEMINI calculations are in good agreement with the experimental data, although the theoretical values are slightly higher than the experimental
data. The minor overestimation may partly be attributed to the slightly higher incident energy in TDHF+GEMINI calculation.
The agreement between theoretical results and experimental data is quite satisfactory, considering that TDHF has no parameter adjusted on the reaction dynamics.
These results give us confidence in providing
rather reliable predictions of transfer dynamics for the production of new proton-rich nuclei in the
vicinity of $^{100}$Sn.

Since the low production cross sections of proton-rich nuclei in $^{100}$Sn region via projectile fragmentation and fusion-evaporation reactions
restrict further experimental exploration toward and beyond the proton drip-line,
multinucleon transfer reaction is proposed as an alternative reaction mechanism for the production of new proton-rich nuclei. 
Figure~\ref{Fig:orientation} shows the production cross sections of proton-rich Sn isotopes using multinucleon transfer (MNT), projectile fragmentation, and fusion-evaporation reactions.
The MNT reaction $^{58}$Ni+$^{112}$Sn is performed with TDHF+GEMINI approach at the energy of $1.4V_\text{B}$ for the tip (open circles) and side (solid circles) collisions, respectively.
The measured cross sections using a fragmentation of $^{124}$Xe projectile on a Be target~\cite{Suzuki2013_NIMB317-756,Celikovic2016_PRL116-162501} (solid diamonds) and fusion-evaporation reaction $^{58}$Ni+$^{50}$Cr~\cite{Karny2005_EPJA25-135} (open diamonds) are included for comparison.
We see that the production cross sections present a rapid decrease with the decrease of neutron number for all three reaction mechanisms.
For the nuclei closer to the stability line, the MNT, fragmentation, and fusion reactions provide the similar cross sections.
However, the cross sections of the most proton-rich Sn isotopes via MNT reaction are found to be several orders of magnitude higher than the other two experimental measurements.
For example, the cross section of $^{100}$Sn nucleus, the most proton-rich Sn isotopes produced so far, is around $10^{-9}$ mb via the fragmentation
of $^{124}$Xe projectile~\cite{Suzuki2013_NIMB317-756}, while MNT reaction predicts the cross section to be four orders of magnitude higher lying
in between $10^{-4}-10^{-6}$ mb. 
The production mechanism of proton-rich Sn isotopes in $^{58}$Ni+$^{112}$Sn is mainly dominated by the neutron transfers due to the stabilizing effect of
closed proton shells of 28 and 50 in projectile and target nuclei.
Our results suggest that new proton-rich Sn isotopes beyond $^{100}$Sn may be expected to be produced via MNT reactions
at the current experimental facilities, which are difficult to be produced by other methods.
These encouraging results clearly reveal the advantage of the proposed MNT reaction
for the production of new proton-rich nuclei in the vicinity of $^{100}$Sn.

\begin{figure*}
\includegraphics[width=0.97\textwidth]{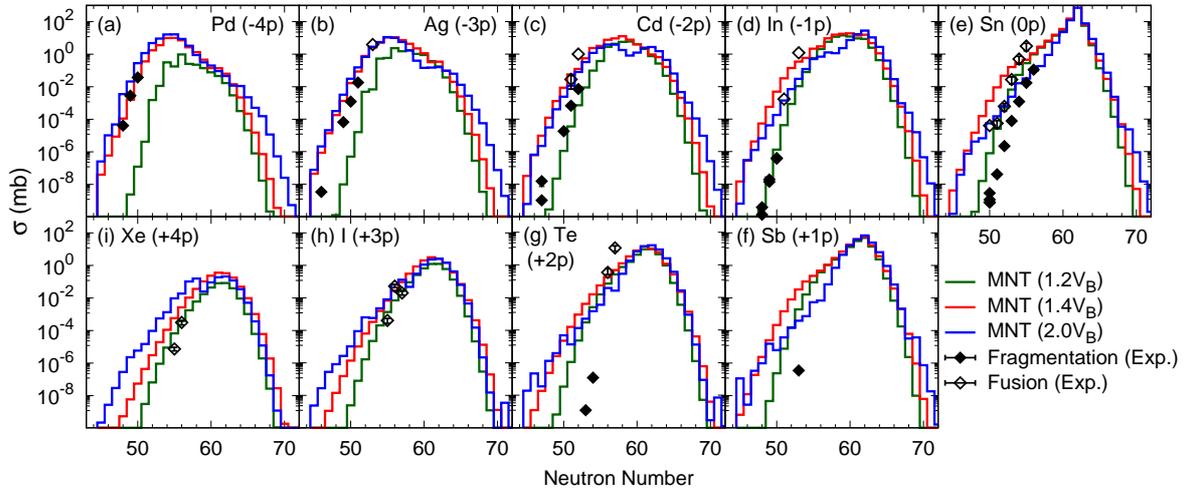}\\
\caption{(Color online) Energy dependence of production cross sections for $Z=46-54$ isotopes. The MNT reaction $^{58}$Ni+$^{112}$Sn is performed with TDHF+GEMINI approach
at center-of-mass energies of 1.2, 1.4, and 2.0$V_\text{B}$ in the tip collision. The measured cross sections in the projectile fragmentation~\cite{Suzuki2013_NIMB317-756,Celikovic2016_PRL116-162501} (solid diamonds) and fusion-evaporation reactions~\cite{Karny2005_EPJA25-135,Korgul2008_PRC77-034301,Commara2000_NPA669-43,Chartier1996_PRL77-2400} (open diamonds) are included for comparison.}
\label{Fig:energy}
\end{figure*}

To reveal the isotopic trend of new isotopes produced in multinucleon transfer reaction, the production cross sections for $Z=46-54$ isotopes are shown in Fig.~\ref{Fig:energy} at center-of-mass energies of 1.2, 1.4, and 2.0$V_\text{B}$ in the tip collision of $^{58}$Ni+$^{112}$Sn.
We observe that the cross sections, in particular for the proton-rich nuclei, are sensitive to the incident energy, while the cross sections of
stable and neutron-rich isotopes weakly depend on the incident energy. The energy-dependent behavior may partly be attributed to the
deexcitation of primary fragments by the evaporation of neutrons. On the other hand, we found that for proton stripping channels ((a)-(d) panels in Fig.~\ref{Fig:energy}) the
production of proton-rich nuclei is dominated by the mechanism of neck formation leading to the transfer of both protons and neutrons in the same direction, while for proton pickup channels ((f)-(i) panels in Fig.~\ref{Fig:energy}) the
charge equilibrium predominates the production of proton-rich nuclei to reduce the asymmetry of neutron-to-proton ratio $N/Z$ between projectile and target nuclei. These two competing mechanisms leading to the production
of $Z=46-54$ proton-rich nuclei have been shown to be related to the incident energy~\cite{Wu2019_PRC100-014612}.
For proton-rich nuclei, the cross sections at the energies of 1.4 and 2.0$V_\text{B}$ are similar, but
several orders of magnitude higher than those at
1.2$V_\text{B}$. This is consistent with the experimental evidence that the optimal incident energy in multinucleon transfer reactions lies in between $6-8.5$ MeV/nucleon.
The systematic studies of energy dependence of MNT dynamics suggest the optimal incident energy to be between 1.4 and 2.0$V_\text{B}$
for the production of proton-rich nuclei in $^{100}$Sn region.

For a systematic comparison among the different reaction mechanisms, the measured cross sections in the projectile fragmentation~\cite{Suzuki2013_NIMB317-756,Celikovic2016_PRL116-162501} (solid diamonds) and fusion-evaporation reactions~\cite{Karny2005_EPJA25-135,Korgul2008_PRC77-034301,Commara2000_NPA669-43,Chartier1996_PRL77-2400} (open diamonds) are also included in Fig.~\ref{Fig:energy}. We see that for the nuclei around the stability line the cross sections with MNT reaction are close to those measured via fragmentation and fusion-evaporation experiments, but
those of proton-rich nuclei via MNT reactions gradually deviate from the measurements and are predicted to be several orders of magnitude higher than those from conventional experimental methods.
For example, the recently measured cross section of $^{97}$In nucleus via the fragmentation of a 345A MeV $^{124}$Xe beam on a Be target is around $10^{-10}$ mb~\cite{Celikovic2016_PRL116-162501} in contrast to $10^{-6}$ mb in multinucleon transfer reaction $^{58}$Ni+$^{112}$Sn.
For Sb isotopes, $^{104}$Sb is the most proton-rich nucleus measured so far and has the cross section of $10^{-7}$ mb in fragmentation reaction~\cite{Suzuki2013_NIMB317-756} as compared to the $10^{-3}$ mb via MNT reaction.
For the more proton-rich Sb isotopes, the recent experimental observation concerning the stability of $^{103}$Sb against proton emission and half-lives~\cite{Suzuki2017_PRC96-034604} contradicts the previous experiment~\cite{Rykaczewski1995_PRC52-R2310}, while the MNT reaction predicts its 
production cross section to be the order of 100 nb.
The $^{104}$Te nucleus has been observed until very recently via fusion-evaporation reaction~\cite{Auranen2018_PRL121-182501} due to a small cross section of less than $10^{-6}$ mb , as compared to the $10^{-4}$ mb via MNT reaction.

\begin{figure*}
\includegraphics[width=0.97\textwidth]{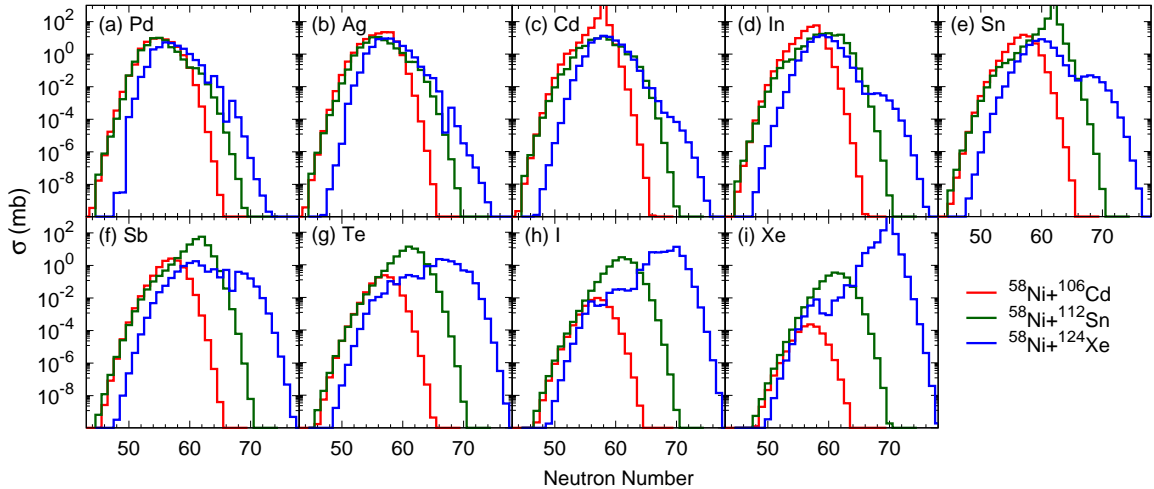}\\
\caption{(Color online) Target-nucleus dependence of production cross sections for $Z=46-54$ isotopes. The MNT reactions of $^{58}$Ni with $^{106}$Cd (red), $^{112}$Sn (green),
and $^{124}$Xe (blue) are performed with TDHF+GEMINI approach at center-of-mass energy of 1.4$V_\text{B}$ in the tip collision.}
\label{Fig:system}
\end{figure*}

Figure~\ref{Fig:system} further clarifies the dependence of production cross sections on reaction systems for $Z=46-54$ isotopes.
The MNT reactions of $^{58}$Ni with $^{106}$Cd (red), $^{112}$Sn (green),
and $^{124}$Xe (blue) are performed with TDHF+GEMINI approach at center-of-mass energy of 1.4$V_\text{B}$ in the tip collision.
The cross sections show complicated and notable dependence on the reaction systems.
For stable and proton-rich nuclei, the cross sections
are quite close for the reactions using $^{106}$Cd and $^{112}$Sn targets, but much smaller cross sections are observed with $^{124}$Xe target 
for the proton-rich nuclei.
For example, the cross sections of $^{100}$Sn nucleus are predicted to be $10^{-3}$ mb with both $^{106}$Cd and $^{112}$Sn targets as compared to
$10^{-7}$ mb with $^{124}$Xe target. On the neutron-rich side, the cross sections exhibit the opposite behavior, i.e, the reaction with $^{124}$Xe target
presents the much larger cross sections than those with $^{106}$Cd and $^{112}$Sn targets.
The remarkable target-nucleus dependence might be attributed to the competition of the two reaction mechanisms
between charge equilibrium and neck formation. In $^{58}$Ni+$^{124}$Xe reaction, the large asymmetry of neutron-to-proton ratio $N/Z$ 
between projectile and target nuclei brings about the suppression of neck formation, leading to the lower cross sections for proton-rich nuclei
and higher ones for neutron-rich nuclei.
Our results suggest that about 19, 13, and 5 new proton-rich isotopes with cross sections of larger than 1 nb may be produced via multinucleon transfer reactions of $^{58}$Ni with $^{112}$Sn, $^{106}$Cd, and $^{124}$Xe, respectively.
These previously unreported proton-rich nuclei with much higher production cross sections arise from the combined effect of several factors of stabilizing effect of shell structure, charge equilibrium, and neck formation.
These results indicate that multinucleon transfer reaction is a more promising way for the production of new proton-rich nuclei near the doubly magic nucleus $^{100}$Sn.

\section{SUMMARY}
\label{summary}

To summarize, multinucleon transfer reactions in low-energy collisions are proposed for the production of new proton-rich nuclei in the vicinity
of $^{100}$Sn, which are of great importance for experimental studies related to astrophysical rp-process and shell evolution far from stability.
The microscopic TDHF+GEMINI results are in good agreement with experimental data without introducing any parameter adjusted for reaction dynamics.
The production cross sections of proton-rich nuclei via multinucleon transfer reactions are found to be
several orders of magnitude higher than those measured in projectile fragmentation and fusion-evaporation reactions.
The large cross sections explicitly reveal the enormous advantage of multinucleon transfer reactions
and provide a good opportunity to discover new proton-rich isotopes in $^{100}$Sn region.
Our results suggest that about 19, 13, and 5 new proton-rich isotopes with cross sections of larger than 1 nb may be produced in multinucleon transfer reactions of $^{58}$Ni with $^{112}$Sn,
$^{106}$Cd, and $^{124}$Xe, respectively.
The reaction mechanisms for the production of these previously unreported nuclei arise from the mixed effects of shell structure, neck evolution, 
and charge equilibrium.
With the development of new experimental techniques as well as the investigation of new reaction mechanisms, the unknown exotic nuclei with extreme neutron-to-proton ratio may become accessible, particularly in the region of the doubly magic $^{100}$Sn nucleus.

\section*{acknowledgments}
This work has been supported by the Strategic Priority Research Program of Chinese Academy of Sciences (Grants Nos. XDB34010000 and XDPB15), the National Natural Science Foundation of China (Grants Nos. 11975237, 11575189, 11790325, 11635003, 11675225, 11961141004, and 11735017), and the National Key R\&D Program of China (Contract No.2018YFA0404402). 
The computations in present work have been performed on the C3S2 computing center in Huzhou University and HPC cluster in Beijing PARATERA Tech Ltd.

\bibliographystyle{elsarticle-num}
\bibliography{ref_arxiv.bib}

\begin{thebibliography}{10}
\expandafter\ifx\csname url\endcsname\relax
  \def\url#1{\texttt{#1}}\fi
\expandafter\ifx\csname urlprefix\endcsname\relax\def\urlprefix{URL }\fi
\expandafter\ifx\csname href\endcsname\relax
  \def\href#1#2{#2} \def\path#1{#1}\fi

\bibitem{Celikovic2016_PRL116-162501}
I.~{\v{C}}elikovi{\'{c}}, M.~Lewitowicz, R.~Gernh\"{a}user, R.~Kr\"{u}cken,
  S.~Nishimura, H.~Sakurai, D.~Ahn, H.~Baba, B.~Blank, A.~Blazhev,
  P.~Boutachkov, F.~Browne, G.~de~France, P.~Doornenbal, T.~Faestermann,
  Y.~Fang, N.~Fukuda, J.~Giovinazzo, N.~Goel, M.~G{\'{o}}rska, S.~Ilieva,
  N.~Inabe, T.~Isobe, A.~Jungclaus, D.~Kameda, Y.-K. Kim, Y.-K. Kwon,
  I.~Kojouharov, T.~Kubo, N.~Kurz, G.~Lorusso, D.~Lubos, K.~Moschner, D.~Murai,
  I.~Nishizuka, J.~Park, Z.~Patel, M.~Rajabali, S.~Rice, H.~Schaffner,
  Y.~Shimizu, L.~Sinclair, P.-A. S\"{o}derstr\"{o}m, K.~Steiger, T.~Sumikama,
  H.~Suzuki, H.~Takeda, Z.~Wang, H.~Watanabe, J.~Wu, Z.~Xu, New isotopes and
  proton emitters{\textendash}crossing the drip line in the vicinity of
  $^{100}${Sn}, Phys. Rev. Lett. 116~(16) (2016) 162501.
\newblock \href {https://doi.org/10.1103/physrevlett.116.162501}
  {\path{doi:10.1103/physrevlett.116.162501}}.

\bibitem{Faestermann2013_PPNP69-85}
T.~Faestermann, M.~G{\'{o}}rska, H.~Grawe, The structure of $^{100}${S}n and
  neighbouring nuclei, Prog. Part. Nucl. Phys. 69 (2013) 85--130.
\newblock \href {https://doi.org/10.1016/j.ppnp.2012.10.002}
  {\path{doi:10.1016/j.ppnp.2012.10.002}}.

\bibitem{Hinke2012_Nature486-341}
C.~B. Hinke, M.~Bohmer, P.~Boutachkov, T.~Faestermann, H.~Geissel, J.~Gerl,
  et~al., Superallowed {G}amow–{T}eller decay of the doubly magic nucleus
  $^{100}$sn, Nature (London) 486 (2012) 341.
\newblock \href {https://doi.org/https://doi.org/10.1038/nature11116}
  {\path{doi:https://doi.org/10.1038/nature11116}}.

\bibitem{Lubos2019_PRL122-222502}
D.~Lubos, J.~Park, T.~Faestermann, R.~Gernh\"auser, R.~Kr\"ucken,
  M.~Lewitowicz, S.~Nishimura, H.~Sakurai, D.~S. Ahn, H.~Baba, B.~Blank,
  A.~Blazhev, P.~Boutachkov, F.~Browne, I.~{\v{C}}elikovi{\'{c}}, G.~de~France,
  P.~Doornenbal, Y.~Fang, N.~Fukuda, J.~Giovinazzo, N.~Goel, M.~G{\'{o}}rska,
  S.~Ilieva, N.~Inabe, T.~Isobe, A.~Jungclaus, D.~Kameda, Y.~K. Kim,
  I.~Kojouharov, T.~Kubo, N.~Kurz, Y.~K. Kwon, G.~Lorusso, K.~Moschner,
  D.~Murai, I.~Nishizuka, Z.~Patel, M.~M. Rajabali, S.~Rice, H.~Schaffner,
  Y.~Shimizu, L.~Sinclair, P.-A. S\"oderstr\"om, K.~Steiger, T.~Sumikama,
  H.~Suzuki, H.~Takeda, Z.~Wang, N.~Warr, H.~Watanabe, J.~Wu, Z.~Xu, Improved
  value for the gamow-teller strength of the $^{100}${S}n beta decay, Phys.
  Rev. Lett. 122~(22) (2019) 222502.
\newblock \href {https://doi.org/10.1103/physrevlett.122.222502}
  {\path{doi:10.1103/physrevlett.122.222502}}.

\bibitem{NaraSingh2011_PRL107-172502}
B.~S. Nara~Singh, Z.~Liu, R.~Wadsworth, H.~Grawe, T.~S. Brock, P.~Boutachkov,
  N.~Braun, A.~Blazhev, M.~G\'orska, S.~Pietri, D.~Rudolph, C.~Domingo-Pardo,
  S.~J. Steer, A.~Ata\c{c}, L.~Bettermann, L.~C\'aceres, K.~Eppinger,
  T.~Engert, T.~Faestermann, F.~Farinon, F.~Finke, K.~Geibel, J.~Gerl,
  R.~Gernh\"auser, N.~Goel, A.~Gottardo, J.~Gr\c{e}bosz, C.~Hinke, R.~Hoischen,
  G.~Ilie, H.~Iwasaki, J.~Jolie, A.~Ka\c{s}ka\c{s}, I.~Kojouharov,
  R.~Kr\"ucken, N.~Kurz, E.~Merch\'an, C.~Nociforo, J.~Nyberg, M.~Pf\"utzner,
  A.~Prochazka, Z.~Podoly\'ak, P.~H. Regan, P.~Reiter, S.~Rinta-Antila,
  C.~Scholl, H.~Schaffner, P.-A. S\"oderstr\"om, N.~Warr, H.~Weick, H.-J.
  Wollersheim, P.~J. Woods, F.~Nowacki, K.~Sieja, ${16}^{+}$ spin-gap isomer in
  $^{96}${Cd}, Phys. Rev. Lett. 107 (2011) 172502.
\newblock \href {https://doi.org/10.1103/PhysRevLett.107.172502}
  {\path{doi:10.1103/PhysRevLett.107.172502}}.

\bibitem{Davies2019_PRC99-021302}
P.~J. Davies, J.~Park, H.~Grawe, R.~Wadsworth, R.~Gernh\"auser, R.~Kr\"ucken,
  F.~Nowacki, D.~S. Ahn, F.~Ameil, H.~Baba, T.~B\"ack, B.~Blank, A.~Blazhev,
  P.~Boutachkov, F.~Browne, I.~{\v{C}}elikovi{\'{c}}, M.~Dewald, P.~Doornenbal,
  T.~Faestermann, Y.~Fang, G.~de~France, N.~Fukuda, A.~Gengelbach, J.~Gerl,
  J.~Giovinazzo, S.~Go, N.~Goel, M.~G{\'{o}}rska, E.~Gregor, H.~Hotaka,
  S.~Ilieva, N.~Inabe, T.~Isobe, D.~G. Jenkins, J.~Jolie, H.~S. Jung,
  A.~Jungclaus, D.~Kameda, G.~D. Kim, Y.-K. Kim, I.~Kojouharov, T.~Kubo,
  N.~Kurz, M.~Lewitowicz, G.~Lorusso, D.~Lubos, L.~Maier, E.~Merchan,
  K.~Moschner, D.~Murai, F.~Naqvi, H.~Nishibata, D.~Nishimura, S.~Nishimura,
  I.~Nishizuka, Z.~Patel, N.~Pietralla, M.~M. Rajabali, S.~Rice, H.~Sakurai,
  H.~Schaffner, Y.~Shimizu, L.~F. Sinclair, P.-A. S\"oderstr\"om, K.~Steiger,
  T.~Sumikama, H.~Suzuki, H.~Takeda, J.~Taprogge, P.~Th\"ole, S.~Valder,
  Z.~Wang, N.~Warr, H.~Watanabe, V.~Werner, J.~Wu, Z.~Y. Xu, A.~Yagi,
  K.~Yoshinaga, Y.~Zhu, Toward the limit of nuclear binding on the ${N}={Z}$
  line: Spectroscopy of $^{96}${C}d, Phys. Rev. C 99~(2) (2019) 021302.
\newblock \href {https://doi.org/10.1103/physrevc.99.021302}
  {\path{doi:10.1103/physrevc.99.021302}}.

\bibitem{Liddick2006_PRL97-082501}
S.~N. Liddick, R.~Grzywacz, C.~Mazzocchi, R.~D. Page, K.~P. Rykaczewski, J.~C.
  Batchelder, C.~R. Bingham, I.~G. Darby, G.~Drafta, C.~Goodin, C.~J. Gross,
  J.~H. Hamilton, A.~A. Hecht, J.~K. Hwang, S.~Ilyushkin, D.~T. Joss,
  A.~Korgul, W.~Kr\'olas, K.~Lagergren, K.~Li, M.~N. Tantawy, J.~Thomson, J.~A.
  Winger, Discovery of $^{109}\mathrm{Xe}$ and $^{105}\mathrm{Te}$:
  Superallowed $\ensuremath{\alpha}$ decay near doubly magic
  $^{100}\mathrm{Sn}$, Phys. Rev. Lett. 97 (2006) 082501.
\newblock \href {https://doi.org/10.1103/PhysRevLett.97.082501}
  {\path{doi:10.1103/PhysRevLett.97.082501}}.

\bibitem{Darby2010_PRL105-162502}
I.~G. Darby, R.~K. Grzywacz, J.~C. Batchelder, C.~R. Bingham, L.~Cartegni,
  C.~J. Gross, M.~Hjorth-Jensen, D.~T. Joss, S.~N. Liddick, W.~Nazarewicz,
  S.~Padgett, R.~D. Page, T.~Papenbrock, M.~M. Rajabali, J.~Rotureau, K.~P.
  Rykaczewski, Orbital dependent nucleonic pairing in the lightest known
  isotopes of {T}in, Phys. Rev. Lett. 105~(16) (2010) 162502.
\newblock \href {https://doi.org/10.1103/physrevlett.105.162502}
  {\path{doi:10.1103/physrevlett.105.162502}}.

\bibitem{Auranen2018_PRL121-182501}
K.~Auranen, D.~Seweryniak, M.~Albers, A.~D. Ayangeakaa, S.~Bottoni, M.~P.
  Carpenter, C.~J. Chiara, P.~Copp, H.~M. David, D.~T. Doherty, J.~Harker,
  C.~R. Hoffman, R.~V.~F. Janssens, T.~L. Khoo, S.~A. Kuvin, T.~Lauritsen,
  G.~Lotay, A.~M. Rogers, J.~Sethi, C.~Scholey, R.~Talwar, W.~B. Walters, P.~J.
  Woods, S.~Zhu, Superallowed $\ensuremath{\alpha}$ decay to doubly magic
  $^{100}\mathrm{Sn}$, Phys. Rev. Lett. 121 (2018) 182501.
\newblock \href {https://doi.org/10.1103/PhysRevLett.121.182501}
  {\path{doi:10.1103/PhysRevLett.121.182501}}.

\bibitem{Barrett2015_PRC91-064615}
J.~S. Barrett, W.~Loveland, R.~Yanez, S.~Zhu, A.~D. Ayangeakaa, M.~P.
  Carpenter, J.~P. Greene, R.~V.~F. Janssens, T.~Lauritsen, E.~A. McCutchan,
  A.~A. Sonzogni, C.~J. Chiara, J.~L. Harker, W.~B. Walters,
  $^{136}\mathrm{Xe}+^{208}\mathrm{Pb}$ reaction: A test of models of
  multinucleon transfer reactions, Phys. Rev. C 91 (2015) 064615.
\newblock \href {https://doi.org/10.1103/PhysRevC.91.064615}
  {\path{doi:10.1103/PhysRevC.91.064615}}.

\bibitem{Vogt2015_PRC92-024619}
A.~Vogt, B.~Birkenbach, P.~Reiter, L.~Corradi,
  T.~Mijatovi\ifmmode~\acute{c}\else \'{c}\fi{}, D.~Montanari, S.~Szilner,
  D.~Bazzacco, M.~Bowry, A.~Bracco, B.~Bruyneel, F.~C.~L. Crespi,
  G.~de~Angelis, P.~D\'esesquelles, J.~Eberth, E.~Farnea, E.~Fioretto,
  A.~Gadea, K.~Geibel, A.~Gengelbach, A.~Giaz, A.~G\"orgen, A.~Gottardo,
  J.~Grebosz, H.~Hess, P.~R. John, J.~Jolie, D.~S. Judson, A.~Jungclaus,
  W.~Korten, S.~Leoni, S.~Lunardi, R.~Menegazzo, D.~Mengoni, C.~Michelagnoli,
  G.~Montagnoli, D.~Napoli, L.~Pellegri, G.~Pollarolo, A.~Pullia, B.~Quintana,
  F.~Radeck, F.~Recchia, D.~Rosso, E.~\ifmmode~\mbox{\c{S}}\else
  \c{S}\fi{}ahin, M.~D. Salsac, F.~Scarlsassara, P.-A. S\"oderstr\"om, A.~M.
  Stefanini, T.~Steinbach, O.~Stezowski, B.~Szpak, C.~Theisen, C.~Ur, J.~J.
  Valiente-Dob\'on, V.~Vandone, A.~Wiens, Light and heavy transfer products in
  $^{136}\mathrm{Xe}+^{238}\mathrm{U}$ multinucleon transfer reactions, Phys.
  Rev. C 92 (2015) 024619.
\newblock \href {https://doi.org/10.1103/PhysRevC.92.024619}
  {\path{doi:10.1103/PhysRevC.92.024619}}.

\bibitem{Watanabe2015_PRL115-172503}
Y.~X. Watanabe, Y.~H. Kim, S.~C. Jeong, Y.~Hirayama, N.~Imai, H.~Ishiyama,
  H.~S. Jung, H.~Miyatake, S.~Choi, J.~S. Song, E.~Clement, G.~de~France,
  A.~Navin, M.~Rejmund, C.~Schmitt, G.~Pollarolo, L.~Corradi, E.~Fioretto,
  D.~Montanari, M.~Niikura, D.~Suzuki, H.~Nishibata, J.~Takatsu, Pathway for
  the production of neutron-rich isotopes around the $\mathrm{N}=126$ shell
  closure, Phys. Rev. Lett. 115 (2015) 172503.
\newblock \href {https://doi.org/10.1103/PhysRevLett.115.172503}
  {\path{doi:10.1103/PhysRevLett.115.172503}}.

\bibitem{Wuenschel2018_PRC97-064602}
S.~Wuenschel, K.~Hagel, M.~Barbui, J.~Gauthier, X.~G. Cao, R.~Wada, E.~J. Kim,
  Z.~Majka, R.~P\l{}aneta, Z.~Sosin, A.~Wieloch, K.~Zelga, S.~Kowalski,
  K.~Schmidt, C.~Ma, G.~Zhang, J.~B. Natowitz, Experimental survey of the
  production of $\ensuremath{\alpha}$-decaying heavy elements in
  $^{238}\mathrm{U}+^{232}\mathrm{Th}$ reactions at 7.5--6.1 {M}e{V}/nucleon,
  Phys. Rev. C 97 (2018) 064602.
\newblock \href {https://doi.org/10.1103/PhysRevC.97.064602}
  {\path{doi:10.1103/PhysRevC.97.064602}}.

\bibitem{Wu2019_PRC100-014612}
Z.~Wu, L.~Guo, Microscopic studies of production cross sections in multinucleon
  transfer reaction $^{58}\mathrm{Ni}+^{124}\mathrm{Sn}$, Phys. Rev. C 100
  (2019) 014612.
\newblock \href {https://doi.org/10.1103/PhysRevC.100.014612}
  {\path{doi:10.1103/PhysRevC.100.014612}}.

\bibitem{Adamian2010_PRC81-057602}
G.~G. Adamian, N.~V. Antonenko, V.~V. Sargsyan, W.~Scheid, Predicted yields of
  new neutron-rich isotopes of nuclei with $\mathrm{Z}=64$--80 in the
  multinucleon transfer reaction $^{48}\mathrm{Ca}+^{238}\mathrm{U}$, Phys.
  Rev. C 81 (2010) 057602.
\newblock \href {https://doi.org/10.1103/PhysRevC.81.057602}
  {\path{doi:10.1103/PhysRevC.81.057602}}.

\bibitem{Wang2012_PRC85-041601}
N.~Wang, E.-G. Zhao, W.~Scheid, S.-G. Zhou, Theoretical study of the synthesis
  of superheavy nuclei with $\mathrm{Z}=119$ and 120 in heavy-ion reactions
  with trans-uranium targets, Phys. Rev. C 85 (2012) 041601.
\newblock \href {https://doi.org/10.1103/PhysRevC.85.041601}
  {\path{doi:10.1103/PhysRevC.85.041601}}.

\bibitem{Zagrebaev2013_PRC87-034608}
V.~I. Zagrebaev, W.~Greiner, Production of heavy trans-target nuclei in
  multinucleon transfer reactions, Phys. Rev. C 87 (2013) 034608.
\newblock \href {https://doi.org/10.1103/PhysRevC.87.034608}
  {\path{doi:10.1103/PhysRevC.87.034608}}.

\bibitem{Wen2013_PRL111-012501}
K.~Wen, F.~Sakata, Z.-X. Li, X.-Z. Wu, Y.-X. Zhang, S.-G. Zhou,
  {N}on-{G}aussian {F}luctuations and {N}on-{M}arkovian {E}ffects in the
  {N}uclear {F}usion {P}rocess: {L}angevin {D}ynamics {E}merging from {Q}uantum
  {M}olecular {D}ynamics {S}imulations, Phys. Rev. Lett. 111~(1) (2013) 012501.
\newblock \href {https://doi.org/10.1103/PhysRevLett.111.012501}
  {\path{doi:10.1103/PhysRevLett.111.012501}}.

\bibitem{Wen2014_PRC90-054613}
K.~Wen, F.~Sakata, Z.-X. Li, X.-Z. Wu, Y.-X. Zhang, S.-G. Zhou, Energy
  dependence of the nucleus-nucleus potential and the friction parameter in
  fusion reactions, Phys. Rev. C 90~(5) (2014) 054613.
\newblock \href {https://doi.org/10.1103/physrevc.90.054613}
  {\path{doi:10.1103/physrevc.90.054613}}.

\bibitem{Zhao2016_PRC94-024601}
K.~Zhao, Z.~Li, Y.~Zhang, N.~Wang, Q.~Li, C.~Shen, Y.~Wang, X.~Wu, Production
  of unknown neutron--rich isotopes in $^{238}\mathrm{U}+{}^{238}\mathrm{U}$
  collisions at near--barrier energy, Phys. Rev. C 94 (2016) 024601.
\newblock \href {https://doi.org/10.1103/PhysRevC.94.024601}
  {\path{doi:10.1103/PhysRevC.94.024601}}.

\bibitem{Bao2016_PRC93-044615}
X.~J. Bao, Y.~Gao, J.~Q. Li, H.~F. Zhang, Possibilities for synthesis of new
  isotopes of superheavy nuclei in cold fusion reactions, Phys. Rev. C 93
  (2016) 044615.
\newblock \href {https://doi.org/10.1103/PhysRevC.93.044615}
  {\path{doi:10.1103/PhysRevC.93.044615}}.

\bibitem{Feng2017_PRC95-024615}
Z.-Q. Feng, Production of neutron-rich isotopes around $\mathrm{N}=126$ in
  multinucleon transfer reactions, Phys. Rev. C 95 (2017) 024615.
\newblock \href {https://doi.org/10.1103/PhysRevC.95.024615}
  {\path{doi:10.1103/PhysRevC.95.024615}}.

\bibitem{Zhu2017_PLB767-437}
L.~Zhu, J.~Su, W.-J. Xie, F.-S. Zhang, Theoretical study on production of heavy
  neutron-rich isotopes around the {N}=126 shell closure in radioactive beam
  induced transfer reactions, Phys. Lett. B 767~(Supplement C) (2017) 437.
\newblock \href
  {https://doi.org/https://doi.org/10.1016/j.physletb.2017.01.082}
  {\path{doi:https://doi.org/10.1016/j.physletb.2017.01.082}}.

\bibitem{Li2018_PLB776-278}
C.~Li, P.~Wen, J.~Li, G.~Zhang, B.~Li, X.~Xu, Z.~Liu, S.~Zhu, F.-S. Zhang,
  Production mechanism of new neutron-rich heavy nuclei in the
  $^{136}\mathrm{Xe}+^{198}\mathrm{Pt}$ reaction, Phys. Lett. B 776 (2018) 278.
\newblock \href
  {https://doi.org/https://doi.org/10.1016/j.physletb.2017.11.060}
  {\path{doi:https://doi.org/10.1016/j.physletb.2017.11.060}}.

\bibitem{Xu2019_CPC43-064105}
X.-X. Xu, G.~Zhang, J.-J. Li, B.~Li, C.~A.~T. Sokhna, X.-R. Zhang, X.-X. Yang,
  S.-H. Cheng, Y.-H. Zhang, Z.-S. Ge, C.~Li, Z.~Liu, F.-S. Zhang, Production of
  exotic neutron-deficient isotopes near {N}, {Z} = 50 in multinucleon transfer
  reactions, Chinese Physics C 43~(6) (2019) 064105.
\newblock \href {https://doi.org/10.1088/1674-1137/43/6/064105}
  {\path{doi:10.1088/1674-1137/43/6/064105}}.

\bibitem{Simenel2012_EPJA48-152}
C.~Simenel, Nuclear quantum many-body dynamics, Eur. Phys. J. A 48~(11) (2012)
  152.
\newblock \href {https://doi.org/10.1140/epja/i2012-12152-0}
  {\path{doi:10.1140/epja/i2012-12152-0}}.

\bibitem{Nakatsukasa2016_RMP88-045004}
T.~Nakatsukasa, K.~Matsuyanagi, M.~Matsuo, K.~Yabana, Time-dependent
  density-functional description of nuclear dynamics, Rev. Mod. Phys. 88 (2016)
  045004.
\newblock \href {https://doi.org/10.1103/RevModPhys.88.045004}
  {\path{doi:10.1103/RevModPhys.88.045004}}.

\bibitem{Simenel2018_PPNP103-19}
C.~Simenel, A.~S. Umar, Heavy-ion collisions and fission dynamics with the
  time--dependent {H}artree-{F}ock theory and its extensions, Prog. Part. Nucl.
  Phys. 103 (2018) 19.
\newblock \href {https://doi.org/https://doi.org/10.1016/j.ppnp.2018.07.002}
  {\path{doi:https://doi.org/10.1016/j.ppnp.2018.07.002}}.

\bibitem{Stevenson2019_PPNP104-142}
P.~D. Stevenson, M.~C. Barton, Low--energy heavy-ion reactions and the {S}kyrme
  effective interaction, Prog. Part. Nucl. Phys. 104 (2019) 142.
\newblock \href {https://doi.org/https://doi.org/10.1016/j.ppnp.2018.09.002}
  {\path{doi:https://doi.org/10.1016/j.ppnp.2018.09.002}}.

\bibitem{Sekizawa2019_FP7-20}
K.~Sekizawa, {TDHF} {T}heory and {I}ts {E}xtensions for the {M}ultinucleon
  {T}ransfer {R}eaction: {A} {M}ini {R}eview, Front. Phys. 7 (2019) 20.
\newblock \href {https://doi.org/10.3389/fphy.2019.00020}
  {\path{doi:10.3389/fphy.2019.00020}}.

\bibitem{GEMINI++_code}
R.~J. Charity, \textcolor[rgb]{0.00,0.00,0.00}{ in \textit{ Joint ICTP-AIEA
  Advanced Work-shop on Model Codes for Spallation Reactions}, Report
  INDC(NDC)-0530 (IAEA, Vienna, 2008), p. 139; The \texttt{GEMINI++} code can
  be downloaded from}\href
  {https://doi.org/https://bitbucket.org/arekfu/gemini}
  {\path{doi:https://bitbucket.org/arekfu/gemini}}.

\bibitem{Charity2010_PRC82-014610}
R.~J. Charity, Systematic description of evaporation spectra for light and
  heavy compound nuclei, Phys. Rev. C 82 (2010) 014610.
\newblock \href {https://doi.org/10.1103/PhysRevC.82.014610}
  {\path{doi:10.1103/PhysRevC.82.014610}}.

\bibitem{Sekizawa2017_PRC96-041601}
K.~Sekizawa, Enhanced nucleon transfer in tip collisions of
  $^{238}\mathrm{U}+^{124}\mathrm{Sn}$, Phys. Rev. C 96 (2017) 041601.
\newblock \href {https://doi.org/10.1103/PhysRevC.96.041601}
  {\path{doi:10.1103/PhysRevC.96.041601}}.

\bibitem{Sekizawa2017_PRC96-014615}
K.~Sekizawa, Microscopic description of production cross sections including
  deexcitation effects, Phys. Rev. C 96 (2017) 014615.
\newblock \href {https://doi.org/10.1103/PhysRevC.96.014615}
  {\path{doi:10.1103/PhysRevC.96.014615}}.

\bibitem{Jiang2018_CPC42-104105}
X.~Jiang, N.~Wang, Production mechanism of neutron-rich nuclei around
  $\mathrm{N}=126$ in the multi-nucleon transfer reaction
  ${}^{132}\mathrm{Sn}+{}^{208}\mathrm{Pb}$, Chin. phys. C 42~(10) (2018)
  104105.
\newblock \href {https://doi.org/10.1088/1674-1137/42/10/104105}
  {\path{doi:10.1088/1674-1137/42/10/104105}}.

\bibitem{Wu2020_SCPMA63-242021}
Z.~Wu, L.~Guo, Production of proton-rich actinide nuclei in the multinucleon
  transfer reaction $^{58}${N}i+$^{232}${T}h, Sci. China-Phys. Mech. Astron.
  63~(4) (2020) 242021.
\newblock \href {https://doi.org/10.1007/s11433-019-1484-0}
  {\path{doi:10.1007/s11433-019-1484-0}}.

\bibitem{Jiang2020_PRC101-014604}
X.~Jiang, N.~Wang, Probing the production mechanism of neutron-rich nuclei in
  multinucleon transfer reactions, Phys. Rev. C 101 (2020) 014604.
\newblock \href {https://doi.org/10.1103/PhysRevC.101.014604}
  {\path{doi:10.1103/PhysRevC.101.014604}}.

\bibitem{Guo2018_PRC98-064609}
L.~Guo, C.~Shen, C.~Yu, Z.~Wu, Isotopic trends of quasifission and
  fusion-fission in the reactions $^{48}\mathrm{Ca}+^{239,244}\mathrm{Pu}$,
  Phys. Rev. C 98 (2018) 064609.
\newblock \href {https://doi.org/10.1103/PhysRevC.98.064609}
  {\path{doi:10.1103/PhysRevC.98.064609}}.

\bibitem{Umar2014_PRC89-034611}
A.~S. Umar, C.~Simenel, V.~E. Oberacker, Energy dependence of potential
  barriers and its effect on fusion cross sections, Phys. Rev. C 89 (2014)
  034611.
\newblock \href {https://doi.org/10.1103/PhysRevC.89.034611}
  {\path{doi:10.1103/PhysRevC.89.034611}}.

\bibitem{Dai2014_PRC90-044609}
G.-F. Dai, L.~Guo, E.-G. Zhao, S.-G. Zhou, Dissipation dynamics and spin-orbit
  force in time-dependent hartree-fock theory, Phys. Rev. C 90 (2014) 044609.
\newblock \href {https://doi.org/10.1103/PhysRevC.90.044609}
  {\path{doi:10.1103/PhysRevC.90.044609}}.

\bibitem{Dai2014_SciChinaPMA57-1618}
G.-F. Dai, L.~Guo, E.-G. Zhao, S.-G. Zhou, Effect of tensor force on
  dissipation dynamics in time-dependent hartree-fock theory, Sci. China-Phys.
  Mech. Astron. 57 (2014) 1618.
\newblock \href {https://doi.org/10.1007/s11433-014-5536-8}
  {\path{doi:10.1007/s11433-014-5536-8}}.

\bibitem{Wang2016_PLB760-236}
N.~Wang, L.~Guo, New neutron-rich isotope production in
  $^{154}\mathrm{Sm}+{}^{160}\mathrm{Gd}$, Phys. Lett. B 760 (2016) 236.
\newblock \href
  {https://doi.org/http://dx.doi.org/10.1016/j.physletb.2016.06.073}
  {\path{doi:http://dx.doi.org/10.1016/j.physletb.2016.06.073}}.

\bibitem{Umar2016_PRC94-024605}
A.~S. Umar, V.~E. Oberacker, C.~Simenel, Fusion and quasifission dynamics in
  the reactions $^{48}\mathrm{Ca}+{}^{249}\mathrm{Bk}$ and
  $^{50}\mathrm{Ti}+{}^{249}\mathrm{Bk}$ using a time-dependent
  {H}artree-{F}ock approach, Phys. Rev. C 94 (2016) 024605.
\newblock \href {https://doi.org/10.1103/PhysRevC.94.024605}
  {\path{doi:10.1103/PhysRevC.94.024605}}.

\bibitem{Stevenson2016_PRC93-054617}
P.~D. Stevenson, E.~B. Suckling, S.~Fracasso, M.~C. Barton, A.~S. Umar, Skyrme
  tensor force in heavy ion collisions, Phys. Rev. C 93 (2016) 054617.
\newblock \href {https://doi.org/10.1103/PhysRevC.93.054617}
  {\path{doi:10.1103/PhysRevC.93.054617}}.

\bibitem{Simenel2017_PRC95-031601}
C.~Simenel, A.~S. Umar, K.~Godbey, M.~Dasgupta, D.~J. Hinde, How the pauli
  exclusion principle affects fusion of atomic nuclei, Phys. Rev. C 95 (2017)
  031601.
\newblock \href {https://doi.org/10.1103/PhysRevC.95.031601}
  {\path{doi:10.1103/PhysRevC.95.031601}}.

\bibitem{Yu2017_SciChinaPMA60-092011}
C.~Yu, L.~Guo, Angular momentum dependence of quasifission dynamics in the
  reaction $^{48}\mathrm{Ca}+^{244}\mathrm{Pu}$, Sci. China-Phys., Mech.
  Astron. 60~(9) (2017) 092011.
\newblock \href {https://doi.org/10.1007/s11433-017-9063-3}
  {\path{doi:10.1007/s11433-017-9063-3}}.

\bibitem{Umar2017_PRC96-024625}
A.~S. Umar, C.~Simenel, W.~Ye, Transport properties of isospin asymmetric
  nuclear matter using the time-dependent hartree-fock method, Phys. Rev. C 96
  (2017) 024625.
\newblock \href {https://doi.org/10.1103/PhysRevC.96.024625}
  {\path{doi:10.1103/PhysRevC.96.024625}}.

\bibitem{Guo2018_PLB782-401}
L.~Guo, C.~Simenel, L.~Shi, C.~Yu, The role of tensor force in heavy-ion fusion
  dynamics, Phys. Lett. B 782 (2018) 401.
\newblock \href
  {https://doi.org/https://doi.org/10.1016/j.physletb.2018.05.066}
  {\path{doi:https://doi.org/10.1016/j.physletb.2018.05.066}}.

\bibitem{Guo2018_PRC98-064607}
L.~Guo, K.~Godbey, A.~S. Umar, Influence of the tensor force on the microscopic
  heavy-ion interaction potential, Phys. Rev. C 98 (2018) 064607.
\newblock \href {https://doi.org/10.1103/PhysRevC.98.064607}
  {\path{doi:10.1103/PhysRevC.98.064607}}.

\bibitem{Scamps2018_Nature564-382}
G.~Scamps, C.~Simenel, Impact of pear-shaped fission fragments on
  mass-asymmetric fission in actinides, Nature 564 (2018) 382--385.
\newblock \href {https://doi.org/10.1038/s41586-018-0780-0}
  {\path{doi:10.1038/s41586-018-0780-0}}.

\bibitem{Sekizawa2019_PRC99-051602}
K.~Sekizawa, K.~Hagino, Time-dependent {H}artree-{F}ock plus {L}angevin
  approach for hot fusion reactions to synthesize the ${Z}=120$ superheavy
  element, Phys. Rev. C 99 (2019) 051602.
\newblock \href {https://doi.org/10.1103/PhysRevC.99.051602}
  {\path{doi:10.1103/PhysRevC.99.051602}}.

\bibitem{Li2019_SciChinaPMA62-122011}
X.~Li, Z.~Wu, L.~Guo, Entrance-channel dynamics in the reaction
  $^{40}${C}a+$^{208}${P}b, Science China Physics, Mechanics {\&} Astronomy
  62~(12) (2019) 122011.
\newblock \href {https://doi.org/10.1007/s11433-019-9435-x}
  {\path{doi:10.1007/s11433-019-9435-x}}.

\bibitem{Godbey2019_PRC100-054612}
K.~Godbey, L.~Guo, A.~S. Umar, Influence of the tensor interaction on heavy-ion
  fusion cross sections, Phys. Rev. C 100 (2019) 054612.
\newblock \href {https://doi.org/10.1103/PhysRevC.100.054612}
  {\path{doi:10.1103/PhysRevC.100.054612}}.

\bibitem{Ayik2019_PRC100-044614}
S.~Ayik, O.~Yilmaz, B.~Yilmaz, A.~S. Umar, Heavy-isotope production in
  $^{136}\mathrm{Xe}+^{208}\mathrm{Pb}$ collisions at
  ${E}_{\mathrm{c}.\mathrm{m}.}=514$ {M}e{V}, Phys. Rev. C 100 (2019) 044614.
\newblock \href {https://doi.org/10.1103/PhysRevC.100.044614}
  {\path{doi:10.1103/PhysRevC.100.044614}}.

\bibitem{Simenel2010_PRL105-192701}
C.~Simenel, Particle transfer reactions with the time-dependent hartree-fock
  theory using a particle number projection technique, Phys. Rev. Lett. 105
  (2010) 192701.
\newblock \href {https://doi.org/10.1103/PhysRevLett.105.192701}
  {\path{doi:10.1103/PhysRevLett.105.192701}}.

\bibitem{Chabanat1998_NPA635-231_NPA643-441}
E.~Chabanat, P.~Bonche, P.~Haensel, J.~Meyer, R.~Schaeffer, {A S}kyrme
  parametrization from subnuclear to neutron star densities {P}art {II}.
  {N}uclei far from stabilities, Nucl. Phys. A 635 (1998) 231; 643 (1998)
  441(E).
\newblock \href {https://doi.org/10.1016/S0375-9474(98)00180-8}
  {\path{doi:10.1016/S0375-9474(98)00180-8}}.

\bibitem{Stevenson2020_IOPSciNotes1-025201}
P.~D. Stevenson, Internuclear potentials from the {S}ky3{D} code, {IOP}
  {SciNotes} 1~(2) (2020) 025201.
\newblock \href {https://doi.org/10.1088/2633-1357/ab952a}
  {\path{doi:10.1088/2633-1357/ab952a}}.

\bibitem{Berg1988_PRC37-178}
A.~M. van~den Berg, W.~Henning, L.~L. Lee, K.~T. Lesko, K.~E. Rehm, J.~P.
  Schiffer, G.~S.~F. Stephans, F.~L.~H. Wolfs, W.~S. Freeman, Quasi-elastic
  processes in $^{58}${N}i $\mathrm{\ensuremath{-}}$ and $^{64}${N}i $\mathrm
  {induced}$ reactions on sn isotopes, Phys. Rev. C 37 (1988) 178--186.
\newblock \href {https://doi.org/10.1103/PhysRevC.37.178}
  {\path{doi:10.1103/PhysRevC.37.178}}.

\bibitem{Suzuki2013_NIMB317-756}
H.~Suzuki, T.~Kubo, N.~Fukuda, N.~Inabe, D.~Kameda, H.~Takeda, K.~Yoshida,
  K.~Kusaka, Y.~Yanagisawa, M.~Ohtake, H.~Sato, Y.~Shimizu, H.~Baba,
  M.~Kurokawa, T.~Ohnishi, K.~Tanaka, O.~Tarasov, D.~Bazin, D.~Morrissey,
  B.~Sherrill, K.~Ieki, D.~Murai, N.~Iwasa, A.~Chiba, Y.~Ohkoda, E.~Ideguchi,
  S.~Go, R.~Yokoyama, T.~Fujii, D.~Nishimura, H.~Nishibata, S.~Momota,
  M.~Lewitowicz, G.~DeFrance, I.~Celikovic, K.~Steiger, Production cross
  section measurements of radioactive isotopes by {BigRIPS} separator at
  {RIKEN} {RI} beam factory, Nucl. Instrum. Methods Phys. Res., Sect. B 317
  (2013) 756--768.
\newblock \href {https://doi.org/10.1016/j.nimb.2013.08.049}
  {\path{doi:10.1016/j.nimb.2013.08.049}}.

\bibitem{Karny2005_EPJA25-135}
M.~Karny, L.~Batist, A.~Banu, F.~Becker, A.~Blazhev, K.~Burkard,
  W.~Br\"{u}chle, J.~D\"{o}ring, T.~Faestermann, M.~G{\'{o}}rska, H.~Grawe,
  Z.~Janas, A.~Jungclaus, M.~Kavatsyuk, O.~Kavatsyuk, R.~Kirchner, M.~L.
  Commara, S.~Mandal, C.~Mazzocchi, K.~Miernik, I.~Mukha, S.~Muralithar,
  C.~Plettner, A.~P{\l}ochocki, E.~Roeckl, M.~Romoli, K.~Rykaczewski,
  M.~Sch\"{a}del, K.~Schmidt, R.~Schwengner, J.~{\.{Z}}ylicz, Beta-decay
  studies near $^{100}${S}n, Eur. Phys. J. A 25~(S1) (2005) 135--138.
\newblock \href {https://doi.org/10.1140/epjad/i2005-06-037-9}
  {\path{doi:10.1140/epjad/i2005-06-037-9}}.

\bibitem{Korgul2008_PRC77-034301}
A.~Korgul, K.~P. Rykaczewski, C.~J. Gross, R.~K. Grzywacz, S.~N. Liddick,
  C.~Mazzocchi, J.~C. Batchelder, C.~R. Bingham, I.~G. Darby, C.~Goodin, J.~H.
  Hamilton, J.~K. Hwang, S.~V. Ilyushkin, W.~Kr{\'{o}}las, J.~A. Winger,
  {Toward} $^{100}${S}n: Studies of excitation functions for the reaction
  between $^{58}${N}i and $^{54}${F}e ions, Phys. Rev. C 77~(3) (2008) 034301.
\newblock \href {https://doi.org/10.1103/physrevc.77.034301}
  {\path{doi:10.1103/physrevc.77.034301}}.

\bibitem{Commara2000_NPA669-43}
M.~L. Commara, J.~G. del Campo, A.~D'Onofrio, A.~Gadea, M.~Glogowski,
  P.~Jarillo-Herrero, N.~Belcari, R.~Borcea, G.~de~Angelis, C.~Fahlander,
  M.~G\'{o}rska, H.~Grawe, M.~Hellstr\"{o}m, R.~Kirchner, M.~Rejmund, V.~Roca,
  E.~Roeckl, M.~Romano, K.~Rykaczewski, K.~Schmidt, F.~Terrasi, Production of
  very neutron-deficient isotopes near sn via reactions involving
  light-particle and cluster emission, Nucl. Phys. A 669~(1-2) (2000) 43--50.
\newblock \href {https://doi.org/10.1016/s0375-9474(99)00814-3}
  {\path{doi:10.1016/s0375-9474(99)00814-3}}.

\bibitem{Chartier1996_PRL77-2400}
M.~Chartier, G.~Auger, W.~Mittig, A.~L\'{e}pine-Szily, L.~K. Fifield, J.~M.
  Casandjian, M.~Chabert, J.~Ferm\'{e}, A.~Gillibert, M.~Lewitowicz, M.~M.
  Cormick, M.~H. Moscatello, O.~H. Odland, N.~A. Orr, G.~Politi, C.~Spitaels,
  A.~C.~C. Villari, Mass measurement of $^{100}$sn, Phys. Rev. Lett. 77~(12)
  (1996) 2400--2403.
\newblock \href {https://doi.org/10.1103/physrevlett.77.2400}
  {\path{doi:10.1103/physrevlett.77.2400}}.

\bibitem{Suzuki2017_PRC96-034604}
H.~Suzuki, T.~Kubo, N.~Fukuda, N.~Inabe, D.~Kameda, H.~Takeda, K.~Yoshida,
  K.~Kusaka, Y.~Yanagisawa, M.~Ohtake, H.~Sato, Y.~Shimizu, H.~Baba,
  M.~Kurokawa, K.~Tanaka, O.~B. Tarasov, D.~Bazin, D.~J. Morrissey, B.~M.
  Sherrill, K.~Ieki, D.~Murai, N.~Iwasa, A.~Chiba, Y.~Ohkoda, E.~Ideguchi,
  S.~Go, R.~Yokoyama, T.~Fujii, D.~Nishimura, H.~Nishibata, S.~Momota,
  M.~Lewitowicz, G.~DeFrance, I.~Celikovic, K.~Steiger, Discovery of new
  isotopes $^{81,82}\mathrm{Mo}$ and $^{85,86}\mathrm{Ru}$ and a determination
  of the particle instability of ${}^{103}${S}b, Phys. Rev. C 96 (2017) 034604.
\newblock \href {https://doi.org/10.1103/PhysRevC.96.034604}
  {\path{doi:10.1103/PhysRevC.96.034604}}.

\bibitem{Rykaczewski1995_PRC52-R2310}
K.~Rykaczewski, R.~Anne, G.~Auger, D.~Bazin, C.~Borcea, V.~Borrel, J.~M. Corre,
  T.~D\"orfler, A.~Fomichov, R.~Grzywacz, D.~Guillemaud-Mueller, R.~Hue,
  M.~Huyse, Z.~Janas, H.~Keller, M.~Lewitowicz, S.~Lukyanov, A.~C. Mueller,
  Y.~Penionzhkevich, M.~Pf\"utzner, F.~Pougheon, M.~G. Saint-Laurent,
  K.~Schmidt, W.~D. Schmidt-Ott, O.~Sorlin, J.~Szerypo, O.~Tarasov, J.~Wauters,
  J.~\ifmmode~\dot{Z}\else \.{Z}\fi{}ylicz, Identification of new nuclei at and
  beyond the proton drip line near the doubly magic nucleus
  $^{100}\mathrm{Sn}$, Phys. Rev. C 52 (1995) R2310--R2313.
\newblock \href {https://doi.org/10.1103/PhysRevC.52.R2310}
  {\path{doi:10.1103/PhysRevC.52.R2310}}.

\end{thebibliography}

\end{document}